\documentclass{ws-ijmpe}
\usepackage{graphicx}  
\usepackage{amsmath, amssymb, amsfonts}  
\providecommand{\draftnote}{}

\usepackage{xspace}
\usepackage{float}
\usepackage{csquotes}
\usepackage[super,compress]{cite}
\begin{document}


\catchline{}{}{}{}{}

\title{Investigation of the onset of deconfinement with the NA61/SHINE experiment}

\author{A. Bazgir (for the NA61/SHINE Collaboration)}

\address{Institute of Physics, Jan Kochanowski University,\\
25-406 Kielce, Poland\\
ali.bazgir@cern.ch}

\maketitle

\begin{abstract}
High-energy heavy-ion collisions provide a unique framework for studying the phase transition of strongly interacting matter. The NA61/SHINE experiment, located in the North Area of CERN's SPS, is a fixed-target facility designed to perform a systematic exploration of the QCD phase diagram. This is achieved through a two-dimensional scan that varies both the beam momentum (from $13$\textit{A} to $150/158$\textit{A} GeV/$c$\xspace) and the size of the colliding systems (\textit{p}+\textit{p}\xspace, \textit{p}+Pb, Be+Be, Ar+Sc, Xe+La, Pb+Pb, O+O). Such a wide scan enables detailed studies of how collision dynamics evolve with system size and energy.
A central objective of the NA61/SHINE research program is to investigate the onset of deconfinement—the transition from hadronic matter to a quark-gluon plasma (QGP)—by analyzing observables such as the strangeness-to-entropy ratio, where entropy is proportional to pion yields. According to the Statistical Model of the Early Stage (SMES), this ratio is expected to exhibit a horn-like structure within the SPS energy range. This article discusses the theoretical framework of the SMES, its assumptions, and compares recent NA61/SHINE results with other experimental data worldwide, contributing to a deeper understanding of the QCD phase transition.
\end{abstract}



\section{Introduction}

The onset of deconfinement refers to the region of collision parameters (e.g., beam energy or thermodynamic parameters) at which a deconfined state of strongly interacting matter begins to form. The existence of this phenomenon was predicted by the Statistical Model of the Early Stage (SMES) \cite{Gazdzicki:1998vd}, which located it in the lower-energy range accessible to the CERN Super Proton Synchrotron (SPS) accelerator. This prediction led to a dedicated energy scan of hadron production in Pb+Pb collisions, carried out by the NA49 experiment at the CERN SPS (the forty-ninth experiment conducted at CERN’s North Area). The results obtained by NA49 confirmed the presence of the onset of deconfinement, as evidenced by the non-monotonic energy dependence of the measures approximately proportional to strangeness-to-entropy ratio \cite{NA49_Pb,Afanasiev:2002}, where entropy was assumed to be proportional to pion yields.

The Statistical Model of the Early Stage (SMES) was developed in the late 1990s based on the first measurements of central Pb+Pb collisions at CERN SPS energies (see Ref.~\cite{Gazdzicki:1998vd}). The model describes the transition between hadronic matter and the QGP and proposes several signals of the onset of deconfinement that can be tested experimentally. In the SMES, only systems produced with all conserved charges equal to zero are considered. Therefore, the properties are fully defined by the available energy and the volume in which production takes place. Only a fraction of the total collision energy is transformed into new degrees of freedom (production of new particles); the rest escapes the fireball in forward and backward directions. The most important postulate of the SMES is that at the early stage of a heavy-ion collision, a maximum-entropy (equilibrium) state is created. The probability of a macroscopic state is proportional to the number of its microscopic realizations.

The SMES predicts that at high collision energies the pure QGP state exists, whereas at low energies only a hadronic gas is present. For intermediate SPS energies, the existence of a mixed phase is postulated. The change from one state to the other should be reflected by a rapid change in the energy dependence of some properties of the produced particles.

The Landau model~\cite{Landau:1953gs} describes the system's expansion process using hydrodynamic equations for a perfect fluid that incorporate the conservation laws. In the model, the entropy and initial temperature of matter are proportional to the $F$\xspace$\left(\sqrt{\text{GeV}}\right)$\xspace, the Fermi-Landau variable:

\begin{equation}
    F=\left[\frac{\left(\sqrt{s_{\text{NN}}}-2m_{\text{N}}\right)^3}{\sqrt{s_{\text{NN}}}}\right]^{1/4} \approx \sqrt[4]{s_{\text{NN}}} ,
\end{equation}
where $\sqrt{s_{\text{NN}}}$ is the collision center-of-mass energy per nucleon pair and $m_\text{N}$ is the nucleon mass. The use of the Fermi-Landau energy variable is convenient for future considerations since the mean multiplicity of pions in nucleon-nucleon~\cite{Abgrall:2013pp_pim} collisions is approximately proportional to $F$.

\begin{figure}[!tb]
    \centering
    
    \begin{tabular}{ccc}
        \includegraphics[width=0.32\textwidth]{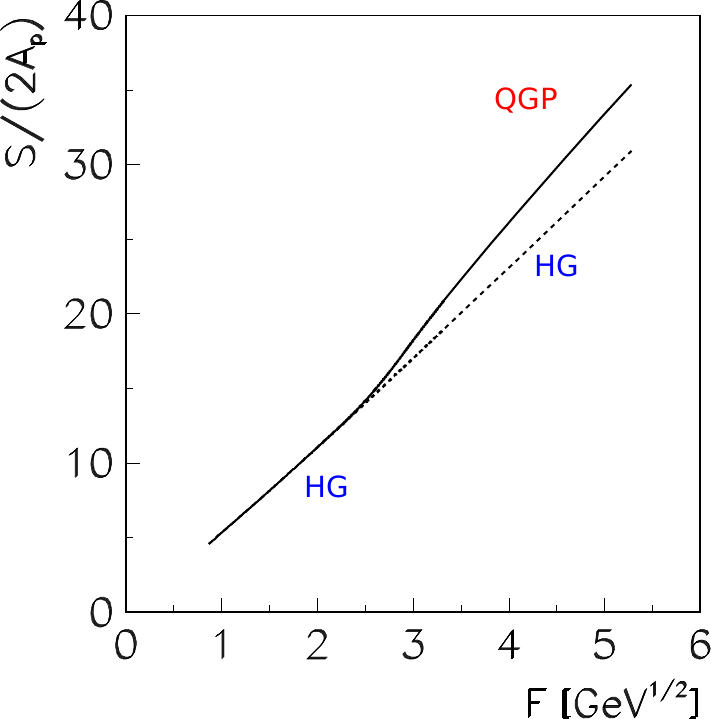} &
        \includegraphics[width=0.32\textwidth]{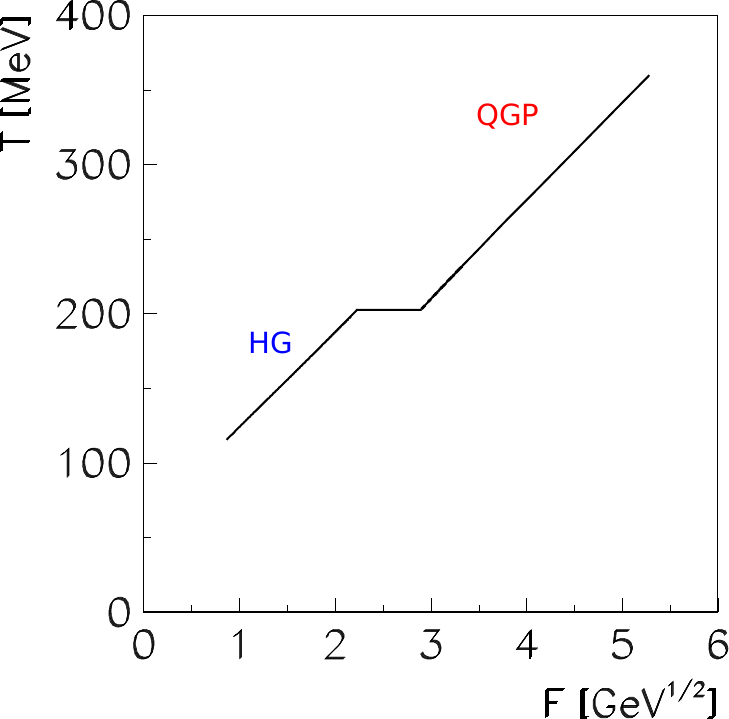} &
        \includegraphics[width=0.32\textwidth]{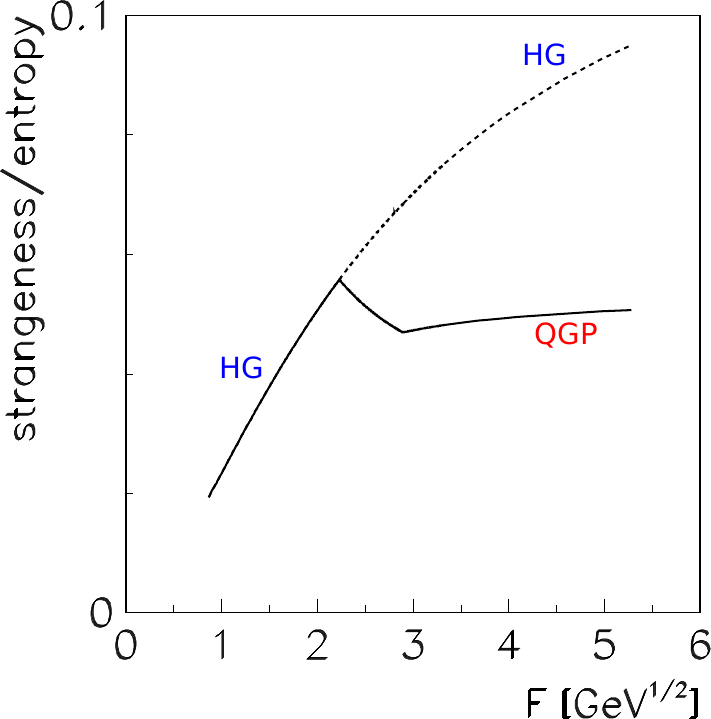}\\
    \end{tabular}
    \caption{Phase transition signatures predicted by the SMES. \textit{ Left}: enhanced entropy production (``\textit{kink}"), \textit{middle}: a plateau in temperature (``\textit{step}"), \textit{right}: non-monotonic strangeness-to-entropy ratio (``\textit{horn}"). All observables are shown as functions of the Fermi-Landau energy measure $F\approx \sqrt[4]{s_\mathrm{NN}}$. Figures taken from Ref.~\cite{Gazdzicki:1998vd}.}
    \label{fig:smes_signatures} 
\end{figure}

The SMES predicts several signatures indicating the onset of deconfinement in collisions involving large $A$+$A$ systems (e.g., Pb+Pb or Au+Au). The proxies of these signatures can be experimentally observed in high-energy heavy-ion collisions. The most significant predictions of SMES are presented in Fig.~\ref{fig:smes_signatures} and discussed in detail below.

\subsection{Enhancement of entropy production: the ``\textit{kink}”}

In heavy-ion collisions, pions are the lightest and most abundantly produced particles and hence carry the majority of the system’s entropy. SMES predicts that entropy production rises with increasing collision energy, but the rate of increase becomes notably higher once the deconfined QGP phase forms, due to its larger number of degrees of freedom.

An experimental proxy for entropy per baryon is the mean pion multiplicity $\langle \pi\rangle$\xspace, normalized by the mean number of wounded nucleons $\langle W\rangle$\xspace. Since the number of wounded nucleons are not measured directly, they must be estimated via theoretical models. According to SMES, the ratio $\langle \pi\rangle/\langle W\rangle$\xspace is sensitive to the state of matter created in the collision, and a change in the slope of its energy dependence signals the onset of deconfinement.

This characteristic behavior, known as the ``\textit{kink}," is typically plotted as a function of the Fermi-Landau energy measure:
\begin{equation}
    \frac{\langle \pi \rangle}{N_\mathrm{part}} \propto g^{1/4} F ,
\end{equation}
where $g$ denotes the number of internal degrees of freedom in a given phase and $N_\mathrm{part}$ denotes the number of participating nucleons. The transition from a hadron gas (HG) to a quark-gluon plasma (QGP) leads to a pronounced increase in $g$, resulting in enhanced entropy production.

\subsection{Plateau in temperature: the ``\textit{step}"}

According to SMES, the system temperature in the early stage of a collision rises with energy until it reaches the transition region. There, a mixed phase of HG and QGP is formed, during which the temperature remains constant. After the transition, temperature increases again in the pure QGP phase. Since direct temperature measurements are not possible, experimentalists use the inverse slope parameter $T$, which is usually extracted from transverse mass ($m_\text{T}$) spectra of kaons:
\begin{equation}
    \frac{\text{d}N}{m_\text{T} \text{d} m_\text{T}}= C\cdot \exp \left( - \frac{m_\text{T}}{T} \right) ,
    \label{eq:exp_mt}
\end{equation}
where $T$ is interpreted as:
\begin{equation}
    T \simeq T_{\text{fo}} + \frac{1}{2} m_0 \beta_\text{T}^2 ,
    \label{eq:inv_sl_thermal}
\end{equation}
with $T_{\text{fo}}$ as the kinetic freeze-out temperature, $m_0$ the particle mass, and $\beta_\text{T}$ the transverse flow velocity. The plateau-like behavior in the energy dependence of $T$—corresponding to the mixed phase—is referred to as the ``\textit{step}."

\subsection{Strangeness-to-entropy ratio: the ``\textit{horn}"}

One of the clearest predicted signals of the deconfinement transition is the energy dependence of the strangeness-to-entropy ratio. In the hadronic phase, the production of strange quarks is suppressed due to the relatively large mass of strange hadrons (e.g., $m_K \approx 500$~MeV), which exceeds the system temperature. In contrast, in the QGP phase, the strange quark mass ($\sim$100~MeV) is lower than the temperature ($T > 150$~MeV), making strangeness production more favorable and resulting in a steady ratio at higher energies.

An observable reflecting this behavior is the $ K^{+}/ \pi^{+}$\xspace multiplicity ratio at mid-rapidity. SMES predicts a sharp peak in this ratio at the onset of deconfinement, after which it stabilizes. This characteristic non-monotonic behavior is called the ``\textit{horn}" and is observed in both Pb+Pb and Au+Au collisions. On the other hand, the data for the $ K^{+}/ \pi^{+}$\xspace multiplicity ratio in \textit{p}+\textit{p}\xspace interactions, reported in Ref.~\cite{Aduszkiewicz:2020}, show no horn-like structure.

\subsection{NA61/SHINE strong interaction program and study the properties of the onset of deconfinement}
Particle production in heavy-ion collisions is a complex, dynamic process that unfolds through multiple stages of fireball evolution. Nevertheless, certain aspects of this process can be effectively described using relatively simple theoretical models. The Hadron Resonance Gas (HRG) model, for example, successfully reproduces particle yields by assuming statistical production from a thermalized source. The SMES offers qualitative predictions for experimental signatures indicative of deconfined matter formation. More sophisticated and comprehensive approaches, such as the Parton--Hadron-String Dynamics (PHSD) model\cite{Bratkovskaya:2011}, also achieve a reasonable description of particle spectra and their anisotropic flow patterns.

Remarkably, all of these models---HRG, SMES, and PHSD---share a significant commonality: they each suggest that the formation of a deconfined phase may already occur in heavy-ion collisions at SPS energies.

In the HRG framework, it was shown that the $ K^{+} / \pi^{+}$\xspace ratio is best reproduced when the $\sigma$-meson and higher-mass resonances are included in the hadronic spectrum~\cite{Andronic:2009}, highlighting the deep connection between this ratio, the hadron mass spectrum, and the QCD phase transition.

The predictions made by SMES are explicitly tied to the presence of a phase transition and are strongly supported by experimental results---particularly the non-monotonic behavior observed in the $ K^{+} / \pi^{+}$\xspace ratio.

Similarly, recent updates to the PHSD model~\cite{Cassing:2016} emphasize that the sharp peak in the $ K^{+} / \pi^{+}$\xspace ratio is best reproduced when both chiral symmetry restoration and deconfinement is incorporated into the modeling framework.

It is important to note, however, that the term ``\textit{phase transition}" is often used interchangeably to refer to both deconfinement and chiral symmetry restoration. A clear distinction between these two phenomena is necessary, as their interrelation---whether they occur simultaneously or separately---remains an open question. Both SMES and PHSD incorporate elements of chiral symmetry restoration and deconfinement in their formulations.


In summary, leading theoretical models offer compelling evidence for the emergence of deconfined matter in heavy-ion collisions at SPS energies. Given that much theoretical and experimental efforts have been devoted to understanding the sharp peak in the $ K^{+} / \pi^{+}$\xspace ratio as a function of collision energy, this observable will also serve as a focal point in the discussion of experimental data from NA61/SHINE.

One of the advantages of using a fixed target in heavy-ion collisions is the ability to access the low transverse-momentum region of phase space and to select events based on the spectator energy of the beam nucleus, independent of the measurements of produced particles. The NA61/SHINE fixed-target experiment~\cite{Abgrall:2014xwa} at CERN focuses on exploring the properties of strongly interacting matter under extreme conditions of temperature and density. One of its central objectives is to investigate the onset of deconfinement—the transition from hadronic matter to a quark-gluon plasma. This transition is predicted by SMES to occur at beam energies accessible at the CERN SPS. NA61/SHINE's goal is to map the phase diagram of strongly interacting matter by identifying signatures of the onset of deconfinement, such as the ``\textit{kink},” ``\textit{horn},” and ``\textit{step}” structures in hadron production observables. The simultaneous observation of the ``\textit{kink},” ``\textit{horn},” and ``\textit{step}” structures in heavy-ion collision data at SPS energies is interpreted as strong evidence for the onset of deconfinement. These structures were first observed in central Pb+Pb collisions~\cite{NA49_Pb,Afanasiev:2002} and are now being studied across different system sizes to determine whether deconfinement can also occur in smaller collision systems~\cite{Abgrall:2014xwa,Aduszkiewicz:2020}. NA61/SHINE’s high-precision data thus provide critical input for understanding the QCD phase transition and the properties of matter created in relativistic heavy-ion collisions.

\section{Spectra analysis methods in NA61/SHINE}

The $tof$--$\mathrm{d}E/\mathrm{d}x$\xspace, $\mathrm{d}E/\mathrm{d}x$\xspace, and \textit{h}$^{-}$\xspace analysis techniques are employed in the NA61/SHINE experiment to investigate charged hadron production. Each method is effective within specific regions of the detector’s acceptance. Importantly, the areas of applicability partially overlap, allowing for cross-validation and complementary coverage of the phase space. This overlap strengthens the reliability and robustness of the final physics results.

The following three analysis methods are used in the study of charged hadron production:

\begin{itemize}
  \item \textbf{The \textit{h}$^{\pmb{-}}$\xspace method:} This approach relies on applying Monte Carlo (MC)-based background subtraction to experimental data. It is specifically designed to extract information on negatively charged pions ($\pi^-$\xspace). This method offers the broadest momentum acceptance among the methods. However, it is associated with relatively large systematic uncertainties—typically around 10\%.

  \item \textbf{The $\pmb{\mathrm{d}E/\mathrm{d}x}$\xspace method:} This technique utilizes measurements of the specific energy loss ($\mathrm{d}E/\mathrm{d}x$\xspace) of charged particles along their trajectories in the Time Projection Chambers (TPCs), combined with their momentum ($p$), to enable particle identification (PID). This allows identification of a range of hadrons beyond $\pi^-$\xspace, including charged kaons ($K^+$), which are especially relevant for investigating the “\textit{horn}” structure—an important indicator of the onset of deconfinement.

  \item \textbf{The $\pmb{tof}$--$\pmb{\mathrm{d}E/\mathrm{d}x}$\xspace method:} This combined method incorporates both d$E$/d$x$ information and time-of-flight ($tof$) measurements, enabling high-quality PID in regions where d$E$/d$x$ performance alone becomes less efficient, particularly for particles with labortory momenta below 5~GeV/$c$. 
\end{itemize}

\subsection{The \textit{h}$^{-}$\xspace method}

In particular, the \textit{h}$^{-}$\xspace method is used in this analysis to determine the production spectra of negatively charged pions ($\pi^-$\xspace). This method is chosen due to its ability to cover the widest range in phase space among the three. The core assumption behind the \textit{h}$^{-}$\xspace method is that the majority of negatively charged hadrons produced in high-energy nuclear collisions are $\pi^-$\xspace mesons. Contamination from other negatively charged particles—such as $K^-$\xspace mesons, antiprotons ($\bar{p}$\xspace), and pions originating from weak decays—is typically small (on the order of 10\%) and can be reliably corrected using detailed Monte Carlo simulations based on event generators.

The analysis proceeds by constructing two-dimensional distributions of these particles as functions of rapidity ($y$\xspace) and transverse momentum ($p_\text{T}$\xspace). These spectra are essential for extracting information about the dynamics of the system created in the collision, such as flow patterns and thermal properties.

Furthermore, the \textit{h}$^{-}$\xspace method allows for a relatively model-independent estimation of pion yields, as it minimizes the need for particle identification cuts or extrapolations in poorly covered acceptance regions. Its effectiveness, combined with robust corrections from MC-based subtraction techniques, makes it particularly valuable for studying central collisions where accurate pion spectra are crucial for understanding the QCD phase diagram and the possible formation of a quark-gluon plasma.

The analysis procedure for the \textit{h}$^{-}$\xspace method briefly consists of the following steps:
\begin{enumerate}
  \item Event and track selection cuts, including the centrality selection.
  \item Determination of the raw double-differential spectra
        $\mathrm{d}^{2}n/\mathrm{d}y\,\mathrm{d}p_{\mathrm{T}}$ of negatively charged hadrons.
  \item Evaluation of MC corrections and calculation of the corrected double-differential spectra $\mathrm{d}^{2}n/\mathrm{d}y\,\mathrm{d}p_{\mathrm{T}}$.
  \item Calculation of the statistical and systematic uncertainties of the corrected
        double-differential spectra $\mathrm{d}^{2}n/\mathrm{d}y\,\mathrm{d}p_{\mathrm{T}}$.
  \item Calculation of the $\mathrm{d}n/\mathrm{d}y$ spectra and the mean multiplicities together with their statistical and systematic uncertainties.
\end{enumerate}

\section{Results}

\begin{figure}[H]
    \centering
    \includegraphics[width=0.49\textwidth]{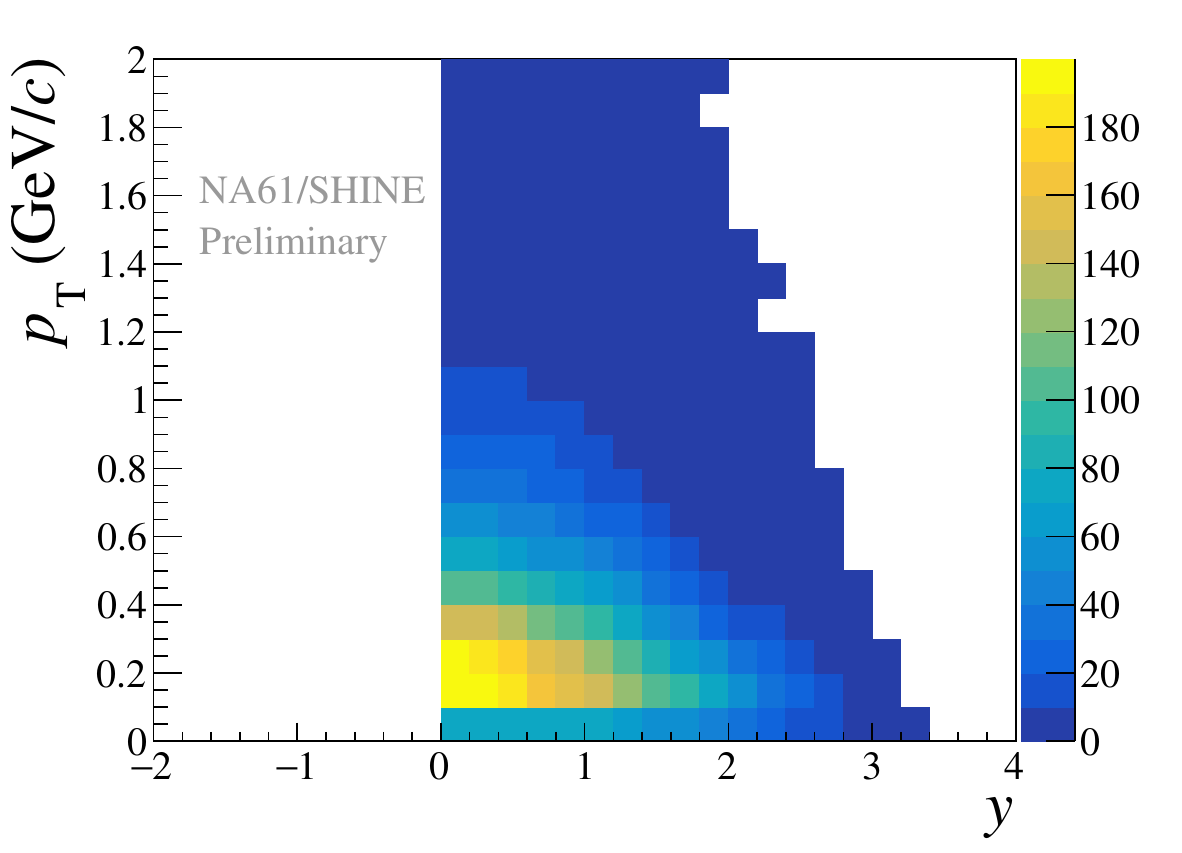}
    \includegraphics[width=0.49\textwidth]{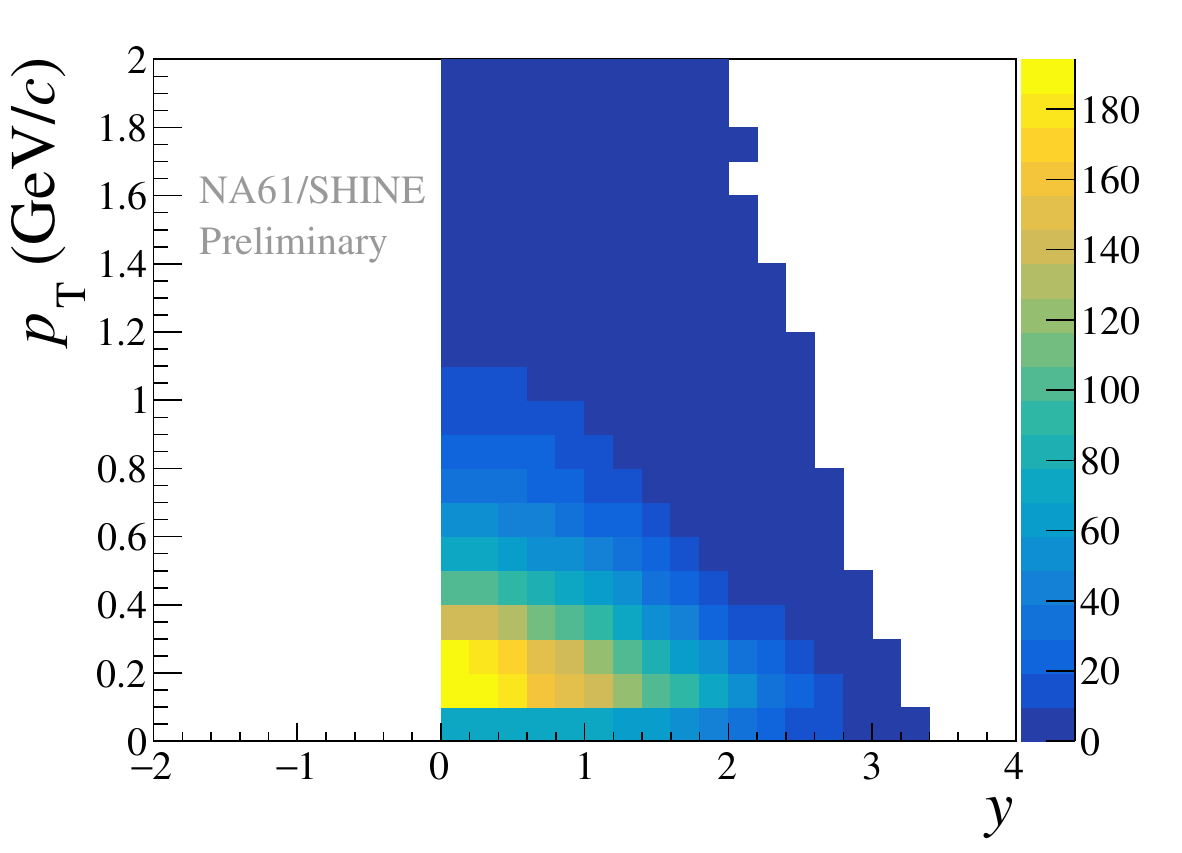}
    \caption{ The $\mathrm{d}^{2}n/\mathrm{d}y\,\mathrm{d}p_{\mathrm{T}}$ spectra of  $\pi^-$\xspace mesons, obtained using the \textit{h}$^{-}$\xspace analysis, measured in $0$--$7.2\%$ (left) and $0$--$10\%$ (right) central Pb+Pb collisions at $30$\textit{A} GeV/$c$\xspace. The rapidity values are defined in the center-of-mass frame of the collision system. Double-differential spectra are normalized using the number of data and MC events.}
    \label{fig:pi-YPT}
\end{figure}

In this section, the preliminary results on $\pi^-$\xspace production in central Pb+Pb collisions at beam momentum of $30$\textit{A} GeV/$c$\xspace are presented. The spectra of $\pi^-$\xspace are obtained using the \textit{h}$^{-}$\xspace analysis technique. The 2D spectra of  $\pi^-$\xspace mesons, after Monte Carlo corrections, are shown in Fig.~\ref{fig:pi-YPT}.

The $\mathrm{d}^{2}n/\mathrm{d}y\,\mathrm{d}p_{\mathrm{T}}$ spectra are projected onto the $p_{\text{T}}$ axis for each $y$ bin separately. These projections are referred to $p_{\text{T}}$ distributions. The extrapolation of the corrected data in transverse momentum is performed independently for each rapidity bin.

The extrapolation of the corrected data in transverse momentum is performed using an exponential formula:

\begin{equation}
  f(p_{\text{T}})= C\,p_{\text{T}} \exp\left(\frac{-(m_\text{T} - m_{\pi^-})}{T}\right) ,
\end{equation}
where $C$ and $T$ denote fit parameters, while $m_{\pi^-}$ and $m_{\text{T}}$ correspond to the charged pion mass and transverse mass, respectively.

\begin{figure}[H]
    \centering
    \includegraphics[width=0.49\textwidth]{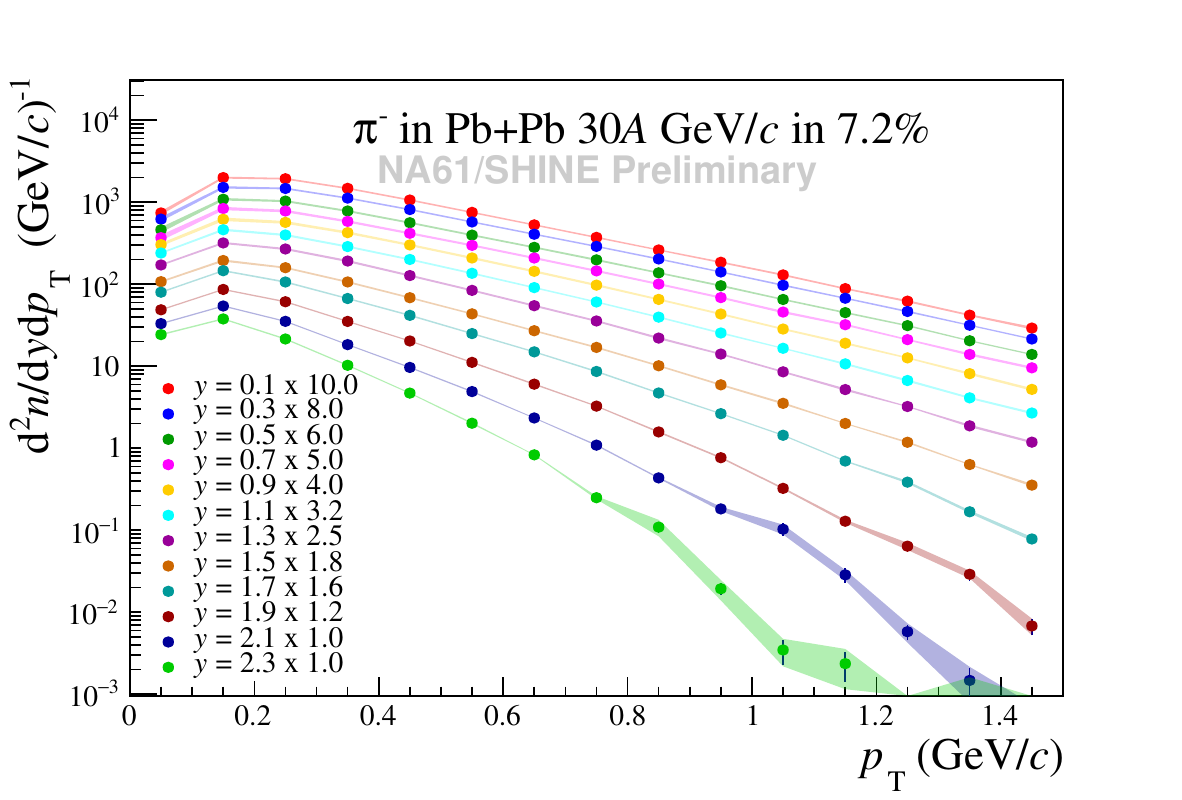}
    \includegraphics[width=0.49\textwidth]{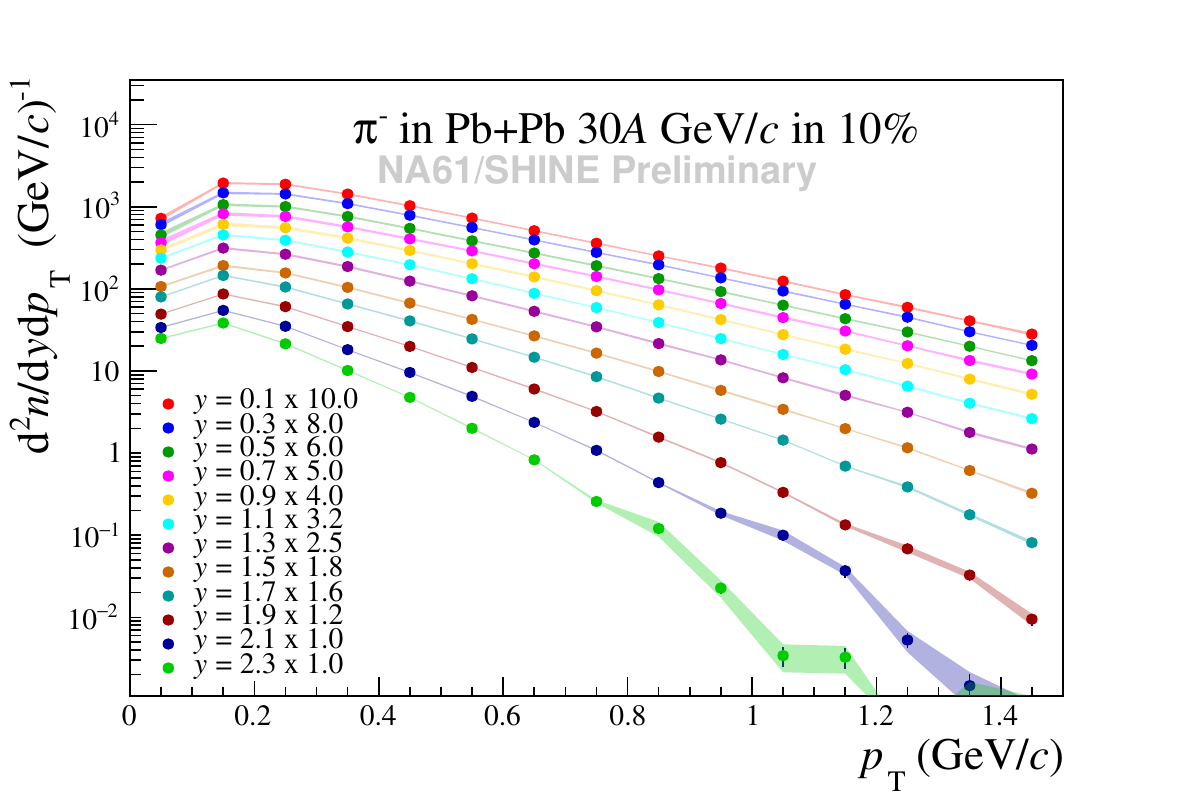}
    \caption{ The $p_{\text{T}}$ spectra of  $\pi^-$\xspace mesons within different rapidity bins, obtained using the \textit{h}$^{-}$\xspace analysis, measured in $0$--$7.2\%$ (left) and $0$--$10\%$ (right) most central Pb+Pb collisions at $30$\textit{A} GeV/$c$\xspace. The rapidity bin width is 0.2, and the rapidity values shown in the legend correspond to the centers of the rapidity bins. Statistical uncertainties are shown by vertical bars, whereas the systematic uncertainties are represented by color bands.}
    \label{fig:pi-PT}
\end{figure}

The $p_{\text{T}}$ distributions of $\pi^-$\xspace  for Pb+Pb collisions for individual $y$\xspace bins are shown in Fig.~\ref{fig:pi-PT} for the most central events of $0$--$7.2\%$ and $0$--$10\%$.

To obtain the rapidity spectra $\text{d}n/\text{d}y$\xspace, the measured $p_{\text{T}}$~bins were summed up to $ p_\text{T}^{\text{max}} = 1.5$\xspace GeV/$c$\xspace  and the integral of the extrapolated function was added:
\begin{equation}
  \frac{\text{d}n}{\text{d}y} =
  \sum_{0}^{p_\text{T}^{\text{max}}}
  \text{d}p_\text{T}
  \left(\frac{\text{d}^{2}n}{\text{d}y\,\text{d}p_\text{T}}\right)_{\text{measured}}
  +
  \int_{p_\text{T}^{\text{max}}}^{\infty}
  f(p_\text{T})\,\text{d}p_\text{T} ,
\end{equation}

\begin{figure}[H]
    \centering
    \includegraphics[width=0.49\textwidth]{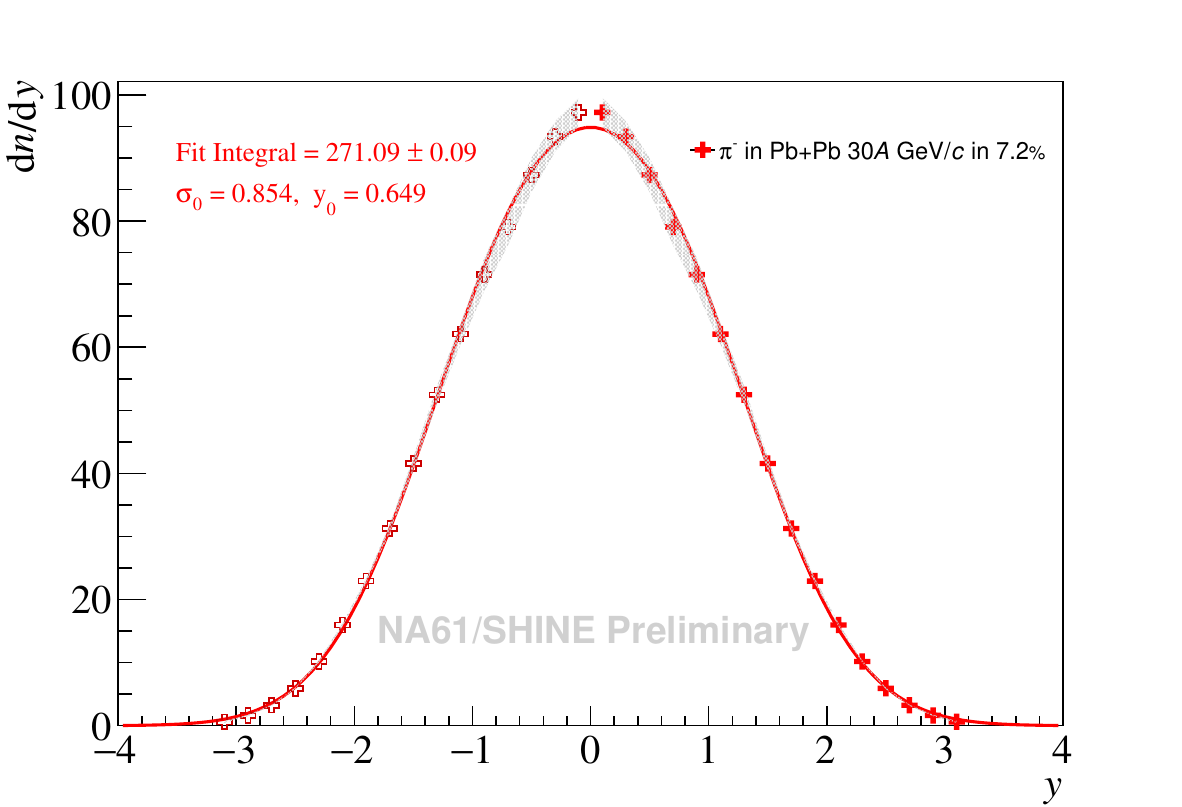}
    \includegraphics[width=0.49\textwidth]{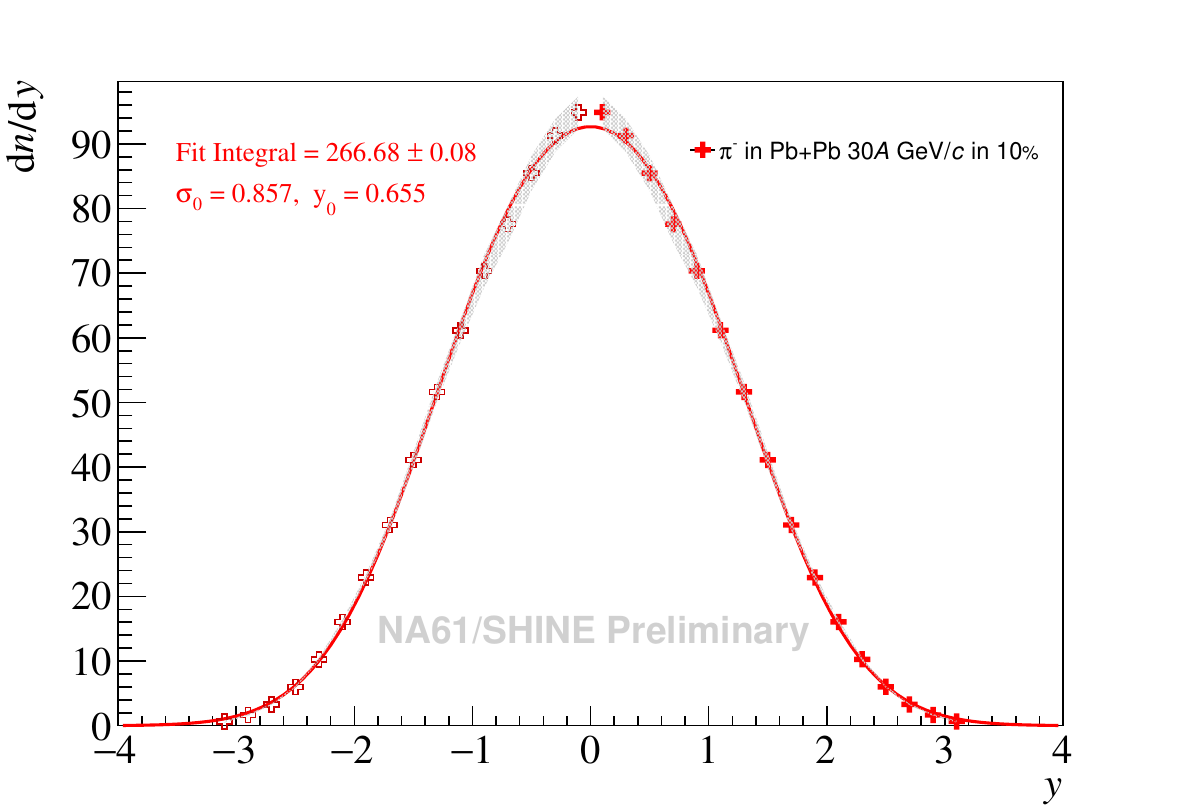}
    \caption{The rapidity distributions of $\pi^{-}$ mesons measured in $0$--$7.2\%$ (left) and $0$--$10\%$ (right) most central Pb+Pb collisions at $30$\textit{A} GeV/$c$\xspace. The statistical uncertainties are smaller than the marker size. The systematic uncertainties are indicated by gray bands.}
    \label{fig:ycorr}
\end{figure}

To extrapolate the rapidity spectra into regions of unmeasured rapidity, a sum of two symmetrically displaced Gaussian functions was employed:

\begin{equation}
    \label{eq:doubleGaussian}
    g(y)=\frac{A_\text{0}}{\sigma_{0}\sqrt{2\pi}}
    \exp\left(-\frac{(y-y_0)^2}{2\sigma_{0}^2}\right)
    +\frac{A_0}{\sigma_{0}\sqrt{2\pi}}
    \exp\left(-\frac{(y+y_0)^2}{2\sigma_{0}^2}\right),
\end{equation}
where $A_0$ denotes the amplitude of the Gaussian, $\sigma_{0}$\xspace is the common standard deviation, and $y_{0}$\xspace represents the displacement of the Gaussian peaks from the center-of-mass rapidity $y=0$. The $\text{d}n/\text{d}y$\xspace distributions of $\pi^-$\xspace mesons, together with the fitted Gaussian distributions, measured in $0$--$7.2\%$ and $0$--$10\%$ central Pb+Pb collisions at $30$\textit{A} GeV/$c$\xspace, are shown in Fig.~\ref{fig:ycorr}.

\begin{figure}[H]
    \centering
    \includegraphics[width=0.49\textwidth]{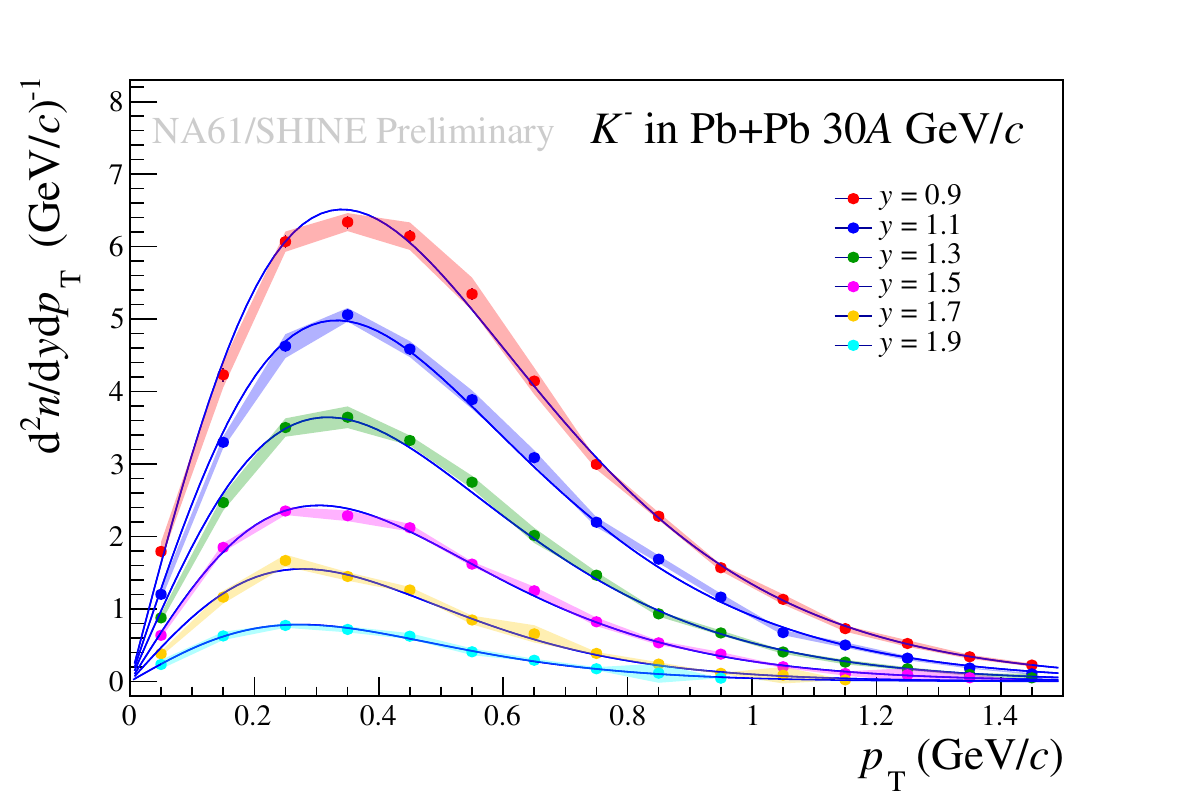}
    \includegraphics[width=0.49\textwidth]{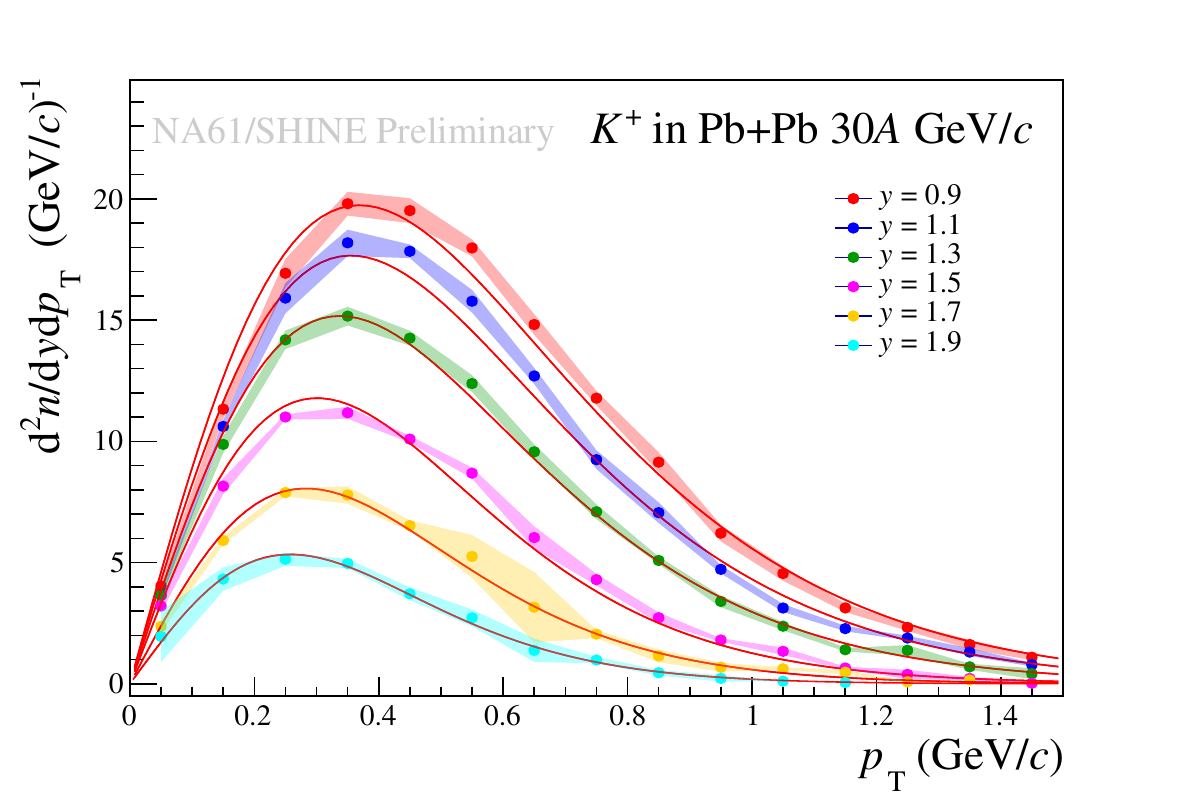}
    \caption{ The $p_{\text{T}}$ spectra of $K^{-}$ (left) and $K^{+}$  (right) mesons, obtained using $\mathrm{d}E/\mathrm{d}x$\xspace method, measured in $0$--$7.2\%$ central Pb+Pb collisions at $30$\textit{A} GeV/$c$\xspace. The rapidity values are defined in the center-of-mass frame of the collision system. The rapidity bin width is 0.2, and the rapidity values shown in the legend correspond to the centers of the rapidity bins. Statistical uncertainties are shown as vertical bars, whereas systematic uncertainties are represented by colored bands.}
    \label{fig:k-k+PT}
\end{figure}

\begin{figure}[H]
    \centering
    \includegraphics[width=0.6\linewidth]{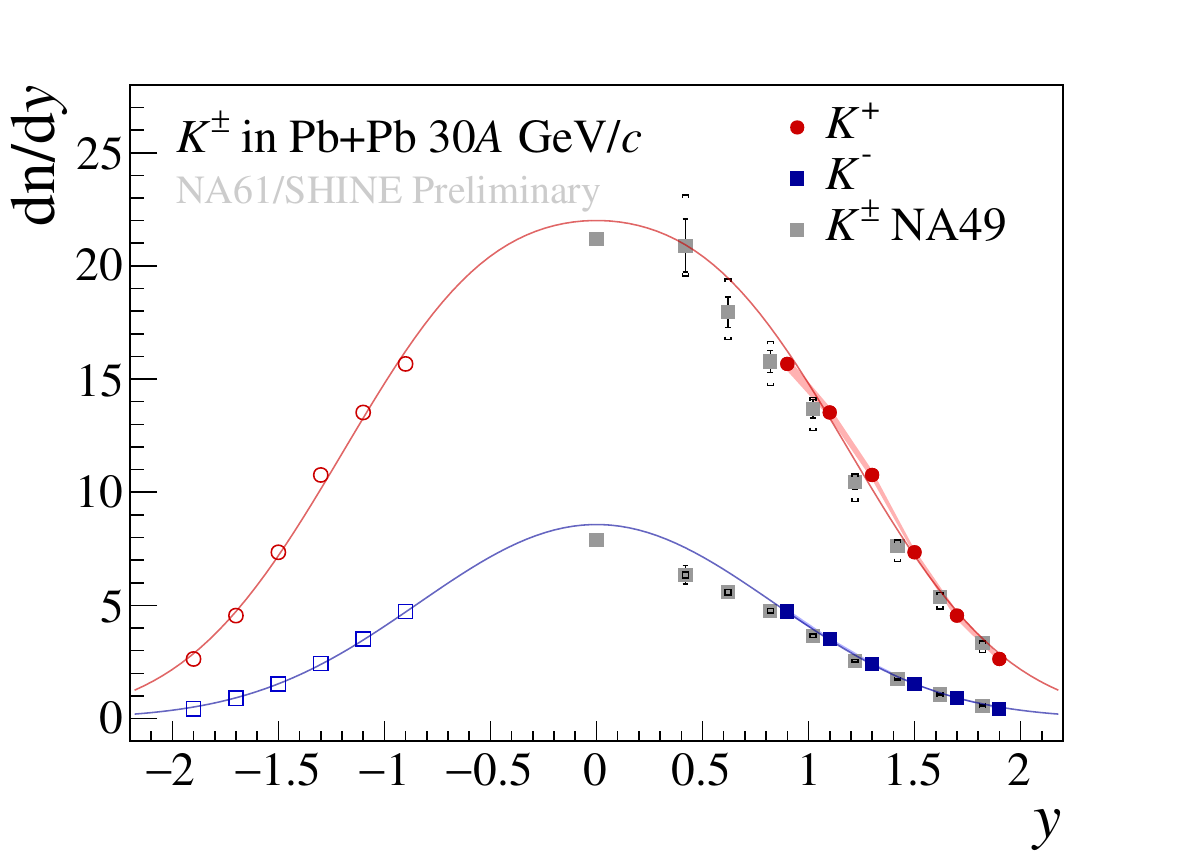}
    \caption{The rapidity distributions of $K^{+}$ and $K^{-}$ mesons, obtained using $\mathrm{d}E/\mathrm{d}x$\xspace method, measured in $0$--$7.2\%$ central Pb+Pb collisions at $30$\textit{A} GeV/$c$\xspace are shown together with their statistical (vertical bars) and systematic (colored bands or brackets) uncertainties. Filled markers represent the directly measured data, while open markers denote the reflections of the data points around mid-rapidity. The NA49 results are taken from Ref.~\cite{NA49_Pb}.}
    \label{fig:ycorr_k-k+}
\end{figure}

The spectra of $K^-$\xspace and $K^+$\xspace, obtained using $\mathrm{d}E/\mathrm{d}x$\xspace method, are corrected for detector acceptance, tracking and reconstruction efficiency, as well as contributions from weak decays and secondary interactions. These corrections were performed using Monte Carlo simulations based on the EPOS1.99~\cite{Werner:EPOS} coupled with GEANT4~\cite{Geant4:Agostinelli}.
The double-differential spectra of identified $K^+$\xspace and $K^-$\xspace mesons produced in the 7.2\% most central Pb+Pb collisions are shown in Fig. \ref{fig:k-k+PT}. Furthermore, the $\text{d}n/\text{d}y$\xspace distributions of $K^+$\xspace and $K^-$\xspace mesons along with comparison with NA49 are shown in Fig. \ref{fig:ycorr_k-k+}.

\section{Onset of deconfinement}

To determine the mean multiplicity $\langle K^{+}\rangle/\langle\pi^{+}\rangle$ ratio, the total yield of $K^{+}$ is obtained by combining the measured $\mathrm{d}n/\mathrm{d}y$ values with the estimated contribution from the unmeasured rapidity regions, evaluated through an extrapolated fit. Within the \textit{h}$^{-}$\xspace analysis method, the $p_{\text{T}}$ spectra of $\pi^-$\xspace mesons were extrapolated into the unmeasured regions and the resulting $\text{d}n/\text{d}y$\xspace spectra were fitted with a sum of two symmetrically displaced Gaussian functions (Eq.~\ref{eq:doubleGaussian}). The mean multiplicity of negatively charged pions, $\langle \pi^-\rangle$\xspace, was then determined from this extrapolated distribution.


At present, no experimental measurements of the mean positively charged pion multiplicity $\langle \pi^+\rangle$\xspace are available for Pb+Pb collisions. Since the Pb+Pb initial state is close to isospin symmetry, the ratio $\langle\pi^{+}\rangle/\langle\pi^{-}\rangle$ is expected to be close to unity. The average multiplicity of $\pi^{+}$ is calculated by scaling the measured $\pi^{-}$ yield using the $\pi^{+}/\pi^{-}$ ratio reported by the \mbox{NA49} experiment~\cite{NA49_Pb}.

\begin{figure}[H]
    \centering
    \includegraphics[width=0.6\linewidth]{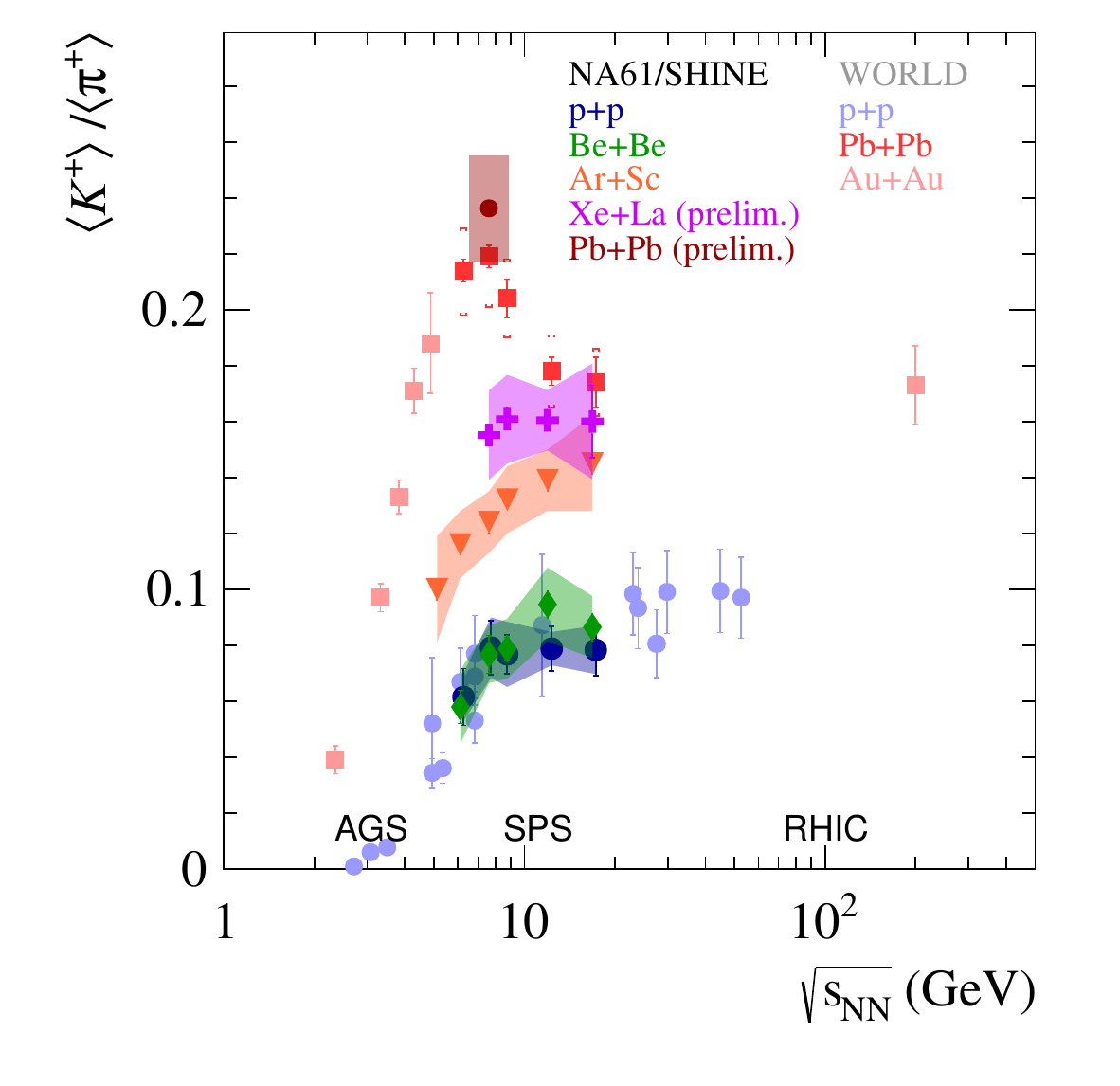}
    \caption{$\langle K^{+}\rangle/\langle\pi^{+}\rangle$ ratio (“\textit{horn}”) as a function of the collision center-of-mass energy per nucleon pair. The references to the published NA61/SHINE results, as well as to the world data, can be found in Ref.~\cite{NA61:2024}.}
    \label{fig:hornresults}
\end{figure}

Figure \ref{fig:hornresults} shows the $\langle K^{+}\rangle/\langle\pi^{+}\rangle$ ratio as a function of the center-of-mass energy. The preliminary results obtained for central Pb+Pb collisions at $30$\textit{A} GeV/$c$\xspace ($\sqrt{s_{\text{NN}}} \approx 7.6$ GeV) are consistent with the NA49 measurements~\cite{NA49_Pb}. The non-monotonic trend observed in the Pb+Pb data indicates a significant modification in the mechanism of strangeness production and is interpreted as the “\textit{horn}” signature associated with the onset of deconfinement.

\section{Summary}

This work presents new results from the NA61/SHINE experiment on central Pb+Pb collisions at a beam momentum of $30A$ GeV/$c$, with a focus on the production of $K^{+}$, $K^{-}$, and $\pi^{-}$ mesons. The $\pi^{-}$ spectra were measured using the $h^{-}$ method. The contamination from other negatively charged particles—such as $K^-$\xspace mesons, antiprotons ($\bar{p}$\xspace), and pions originating from weak decays can be reliably corrected using detailed Monte Carlo simulations based on event generators. The transverse momentum and rapidity spectra of negatively charged pions were described using exponential and Gaussian fits, respectively, which enabled the determination of total particle yields.

To determine the yields of $K^{+}$ and $K^{-}$, particle identification was performed using the specific energy loss ($\mathrm{d}E/\mathrm{d}x$) technique. The $\langle K^{+}\rangle/\langle\pi^{+}\rangle$ ratio was obtained by combining measured and extrapolated spectra, while the $\pi^{+}$ multiplicities were estimated using the $\pi^{+}/\pi^{-}$ ratio measured by the NA49 experiment. The NA61/SHINE result at $\sqrt{s_{\text{NN}}} \approx 7.6$ GeV is consistent with the corresponding NA49 measurement.

The energy dependence of the $\langle K^{+}\rangle/\langle\pi^{+}\rangle$ ratio exhibits a pronounced horn-like structure, commonly interpreted as a signature of the onset of deconfinement. This observation provides supporting evidence for the occurrence of a deconfinement phase transition in this energy region.

\section*{Acknowledgements}
Financial support from the project ``Internationalisation of the Doctoral School of the Jan Kochanowski University'' of the Polish National Agency for Academic Exchange (NAWA STER no. BPI/STE/2023/1/00014) is acknowledged.

\end{document}